\documentclass[sn-apa]{sn-jnl}% APA Reference Style 
\usepackage{graphicx}%
\usepackage{multirow}%
\usepackage{amsmath,amssymb,amsfonts}%
\usepackage{amsthm}%
\usepackage{mathrsfs}%
\usepackage[title]{appendix}%
\usepackage{xcolor}%
\usepackage{textcomp}%
\usepackage{manyfoot}%
\usepackage{booktabs}%
\usepackage{algorithm}%
\usepackage{algorithmicx}%
\usepackage{algpseudocode}%
\usepackage{listings}%
\usepackage{array}
\usepackage{arydshln}
\usepackage{tabu}

\usepackage{multirow}
\usepackage[normalem]{ulem}
\useunder{\uline}{\ul}{}

\usepackage{lineno}
\usepackage{subcaption}
\usepackage{float}
\usepackage{wasysym}

\usepackage{listings}
\usepackage{xcolor}

\definecolor{codegreen}{rgb}{0,0.6,0}
\definecolor{codegray}{rgb}{0.5,0.5,0.5}
\definecolor{codepurple}{rgb}{0.58,0,0.82}
\definecolor{backcolour}{rgb}{0.95,0.95,0.92}

\lstdefinestyle{mystyle}{
    backgroundcolor=\color{backcolour},   
    commentstyle=\color{codegreen},
    keywordstyle=\color{magenta},
    numberstyle=\tiny\color{codegray},
    stringstyle=\color{codepurple},
    basicstyle=\ttfamily\footnotesize,
    breakatwhitespace=false,         
    breaklines=true,                 
    captionpos=b,                    
    keepspaces=true,                 
    numbers=left,                    
    numbersep=5pt,                  
    showspaces=false,                
    showstringspaces=false,
    showtabs=false,                  
    tabsize=2
}

\begin{document}
\title[A Semi-automated Approach to Correcting Eye Tracking Data in Reading Tasks]{Combining Automation and Expertise: A Semi-automated Approach to Correcting Eye Tracking Data in Reading Tasks}


\author[1]{Naser Al Madi}
\email{nsalmadi@colby.edu}
% \affiliation{
%     \institution{Colby College}
%     \city{Waterville}
%     \state{Maine}
%     \country{USA}
% }
% \author[1]{\fnm{Naser} \sur{Al Madi}}\email{nsalmadi@Colby.edu}
% \author[1]{\fnm{Brett} \sur{Torra}}
% \author[1]{\fnm{Yixin} \sur{Li}}
% \author[1]{\fnm{Najam} \sur{Tariq}}
\author[1]{Brett Torra}
\author[1]{Yixin Li}
\author[1]{Najam Tariq}

\affil[1]{\orgdiv{Department of Computer Science}, \orgname{Colby College}, \orgaddress{\city{Waterville}, \state{ME}, \country{USA}}}


\abstract{
In reading tasks drift can move fixations from one word to another or even another line, invalidating the eye tracking recording.  Manual correction is time-consuming and subjective, while automated correction is fast yet limited in accuracy. In this paper we present Fix8 (Fixate), an open-source GUI tool that offers a novel semi-automated correction approach for eye tracking data in reading tasks.  The proposed approach allows the user to collaborate with an algorithm to produce accurate corrections faster without sacrificing accuracy. Through a usability study (N=14) we assess the time benefits of the proposed technique, and measure the correction accuracy in comparison to manual correction. In addition, we assess subjective workload through NASA Task Load Index, and user opinions through Likert-scale questions. Our results show that on average the proposed technique was 44\% faster than manual correction without any sacrifice in accuracy. In addition, users reported a preference for the proposed technique, lower workload, and higher perceived performance compared to manual correction. Fix8 is a valuable tool that offers useful features for generating synthetic eye tracking data, visualization, filters, data converters, and eye movement analysis in addition to the main contribution in data correction.
}




\maketitle
\section{Introduction}\label{sec1}
% \linenumbers

\begin{figure}[h]
    \centering
    \includegraphics[width=0.5\linewidth]{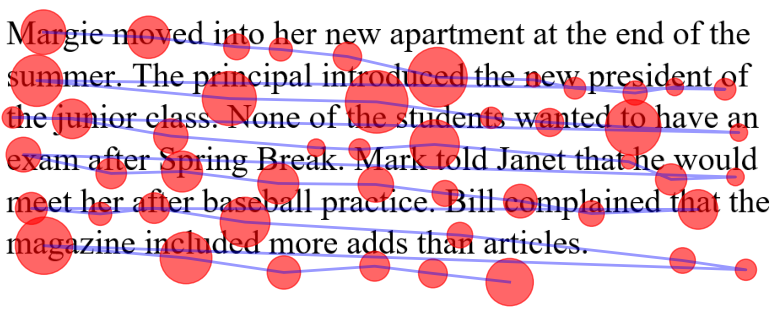}
    \caption{An illustration of drift where fixations deviate from their real position.}
    \label{fig:drift-example}
\end{figure}

Eye tracking has been an important tool in studying the connection between eye movement and human cognition during reading \citep{rayner1998eye}.  Despite the increase in eye tracking accuracy over the years, drift remains a problem in eye tracking experiments.  Drift describes a form of systematic error that deviates the fixation position from its real position \citep{mishra2012heuristic}, this error is often attributed to degradation in calibration, head movement, or changes in lighting conditions \citep{yamaya2015dynamic, carl2013dynamic}.  While a small magnitude of drift is acceptable in video or image viewing experiments, in reading research drift can move fixations to another word or line, as seen in Figure \ref{fig:drift-example}.  This can be problematic considering that many reading research experiments focus on word-level effects, in fact \cite{reichle2015using} found that such systematic error can falsely produce parafoveal-on-foveal effects and spillover effects in reading.  Therefore eye tracking data correction is vital in reading research, and manual and automated methods have been proposed to solve this problem.

Several tools offer the ability to manually correct eye tracking data, such as EyeLink Data Viewer\footnote{https://www.sr-research.com/data-viewer/} and EyeDoctor\footnote{https://websites.umass.edu/eyelab/software/}.  This is often performed by dragging and dropping individual or groups of fixations to a likely position determined by a human expert.  The human corrector is often guided by fixation order, duration, and position in determining the corrected fixation position.  Yet this process is highly subjective, and more importantly laborious and time consuming \citep{cohen2013software, carr2022algorithms, mishra2012heuristic}.  To address the issue of subjectivity, multiple human correctors are often tasked with collaboratively correcting the data, making the process more labor intensive.

Many automated correction algorithms have been proposed to make the correction process faster, less subjective, and easier \citep{abdulin2015person, beymer2005wide, carl2013dynamic, lima2016vertical, martinez2012image, yamaya2017dynamic, zhang2011mode, john2012entropy, carr2022algorithms, al2024advancing}, yet automated algorithms remain less accurate than manual correction \citep{carr2022algorithms}.  In other words, automated algorithms offer a trade-off between speed and accuracy by making the correction process significantly faster in exchange for some reduction in correction accuracy.  While some research studies can tolerate lower accuracy, it is often the case that high accuracy is vital for many research studies.

In this paper, we present a semi-automated correction approach that combines the benefits of manual correction with some of the speed and objectivity benefits of automated algorithms.  The proposed assisted approach allows the user to collaborate with the algorithm to produce accurate corrections in less time with lower workload.  The approach utilizes automated algorithms internally to produce correction suggestions to the user, allowing the user to accept each suggestion or intervene to manually correct a given fixation.  Internally, the technique takes in the human intervention to update future suggestions, making future suggestions more accurate.

The proposed approach is implemented in an open source GUI tool that we name Fix8 (Fixate), which implements several useful features for eye tracking in reading tasks.  In addition to the assisted correction approach, Fix8 offers fully manual and fully automated correction, interactive visualizations of eye tracking data, dynamically generated Areas-of-Interest (AOIs), synthetic data and distortion generation, data filters, hit-test and metric analyses, and data converters.  This tool builds on previous research by incorporating previously implemented algorithms from  \cite{carr2022algorithms}, \cite{al2024advancing}, and data correction techniques that were reviewed by \cite{eskenazi2023best}.

Through a usability study with 14 participants, we assess the time benefits of the proposed assisted technique, and measure the correction accuracy in comparison to manual correction.  In addition, we assess subjective workload through NASA Task Load Index (NASA-TLX) and user opinions through Likert-scale questions.  Our results show that the proposed assisted technique was on average 44\% faster than manual correction without any sacrifice in accuracy.  In addition, users reported a preference for the assisted technique compared to manual correction, lower workload, and higher perceived performance when using the assisted approach.  We make the tool and a collection of datasets publicly available as an open source tool, along with a video illustrating the main features of Fix8.  The code repository, datasets, and video can be found at the following URL: \url{https://github.com/nalmadi/fix8}

This paper is organized as follows: Section \ref{sec2} presents the details of the proposed approach and its internals; Section \ref{sec3} presents a detailed review of the features offered by Fix8; Section \ref{sec4} presents the design, procedure, and results of our usability study; Section \ref{sec5} discusses the findings of our experiments in the context of the proposed tool and our initial objectives; Section \ref{sec6}, presents the conclusion of the present work.

\section{Proposed Approach}\label{sec2}
Fix8 includes useful features for eye tracking data visualization, correction, and analyses, yet the main novel feature is a semi-automated (assisted) approach for correcting eye tracking data in reading tasks.  In this section, we will focus on the assisted correction approach and how it compares to manual and automatic correction.

The proposed approach employs four usability ideas that shape the assisted correction workflow and aid users in correcting eye tracking data. By utilizing visualization, workflow, ergonomics, and human-machine collaboration we aim to make eye tracking data correction faster and reduce user workload without sacrificing accuracy.

% visualization
\textbf{Visualization.} In the proposed approach, fixations are assessed one at a time in a chronological order.  This allows the user to view previous and remaining fixations to make a decision on the position of the current fixation being considered.  Human correctors often rely on previous fixations in making correction decisions utilizing the sequential nature of reading behavior.  Our approach give users the ability to see previous and remaining fixations while also being able to move a progress bar to make more informed correction decisions through an interactive interface.

\begin{figure}[h]%

    \centering

    \begin{subfigure}{\textwidth}
        \centering
        \includegraphics[scale=.45]{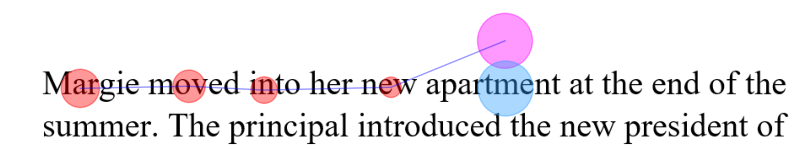}
        \caption{Example of a good suggestion, the fixation is likely on the word ``apartment."}
    \end{subfigure}

    \begin{subfigure}{\textwidth}
        \centering
        \includegraphics[scale=.45]{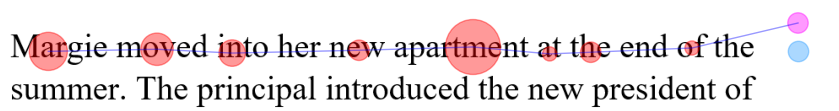}
        \caption{Example of a bad suggestion, the fixation is likely on the word ``the" not the empty space next to it.}
    \end{subfigure}

    \caption{Examples of good and bad suggestions in the assisted correction approach. The magenta circle represents the current fixation being considered, fixations in red are previous fixations, and the blue circle is the suggested correction position.  }
    \label{fig:suggestion_examples}
\end{figure}

% workflow
\textbf{Workflow.} The proposed approach offers a suggested correction from an automated algorithm at each fixation.  As seen in Figure \ref{fig:suggestion_examples} (a), the current fixation is in magenta and the suggested correction is in blue, allowing the user to quickly make a decision to accept or reject the suggestion. This workflow directs the attention of the user to a single correction decision, utilizing a color pattern that clearly highlights the fixation being considered and the suggested correction position.  The user is able to choose one of 13 automated algorithm being used to make the suggestions, these algorithms were implemented by \cite{carr2022algorithms} and \cite{al2024advancing} and we use the algorithm names proposed by \cite{carr2022algorithms}.

% ergonomics
\textbf{Ergonomics.} By utilizing keyboard shortcuts, the user can quickly vet and correct fixations instead of dragging and dropping each fixation manually using the mouse.  The user can press the space-bar to accept the current suggestion, or use the mouse to drag and drop the fixation on the correct position when the user does not agree with the suggestion.  Also, the `a' and `z' buttons can be used to quickly assign the current fixation to the line above it or line bellow it accordingly. Users can place one hand on the keyboard with three fingers on the `a', `z', and space-bar, and the other hand on the mouse to take full advantage of this ergonomic design. This process is faster than manual correction, since users mostly vet suggestions instead of manually dragging-and-dropping all of them, also the suggestion minimizes the cognitive workload and increases objectivity.

% collaboration
\textbf{Collaboration.} The proposed approach relies on the concept of human-machine collaboration to reduce workload and make the correction process faster and less subjective. In Figure \ref{fig:suggestion_examples} (a), the suggested correction (blue circle) is on the word ``apartment'', and the user can press the space-bar to accept this suggestion and move to the next fixation.  This is a good suggestion as it represents a typical drift example where the fixation is slightly above the intended word.  On the other hand, Figure \ref{fig:suggestion_examples} (b) shows an example of an occasional bad suggestion where the suggested correction (in blue) is not on any words.  It is likely that this fixation is on the word ``the", which the user can confirm by exploring the fixations after the current fixation.  In this case, the user can drag and drop the current fixation to the word ``the".  

\begin{figure}[h]
\centering
  \begin{subfigure}{\textwidth}
    \centering
    \includegraphics[scale=.24]{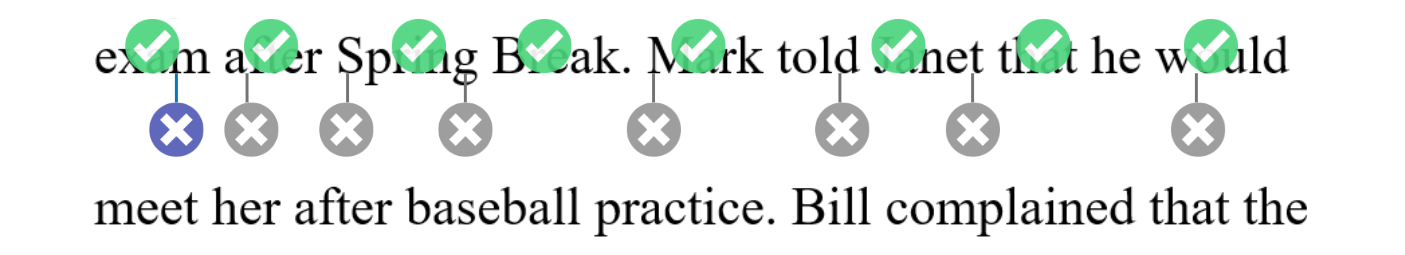}
    \caption{Blue suggestion and remaining suggestions are incorrectly assigned to line above.}
    \label{fig:situation1}
  \end{subfigure}

  \begin{subfigure}{\textwidth}
    \centering
    \includegraphics[scale=.24]{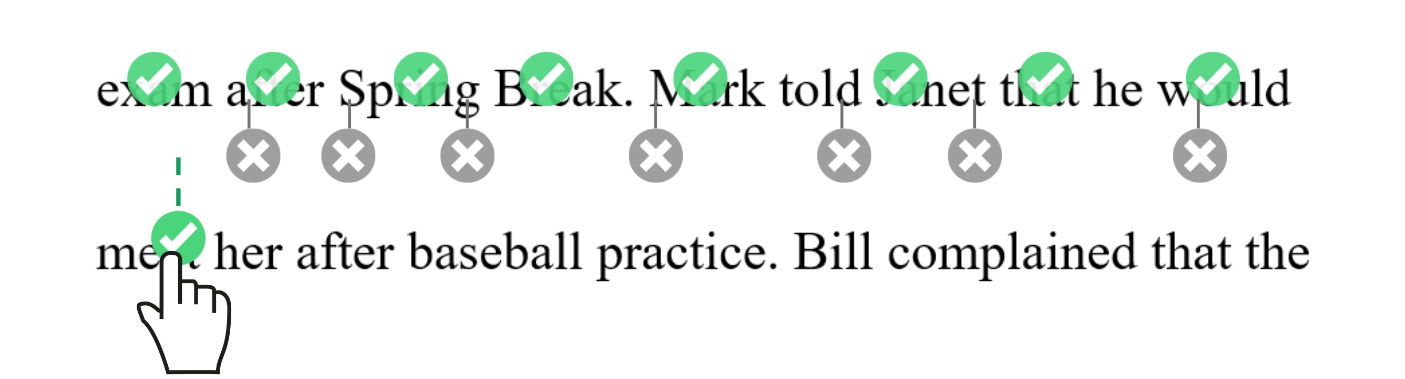}
    \caption{User intervenes by dragging and dropping a fixation to correct it, instead of accepting the suggestion.}
    \label{fig:situation2}
  \end{subfigure}

  \begin{subfigure}{\textwidth}
    \centering
    \includegraphics[scale=.24]{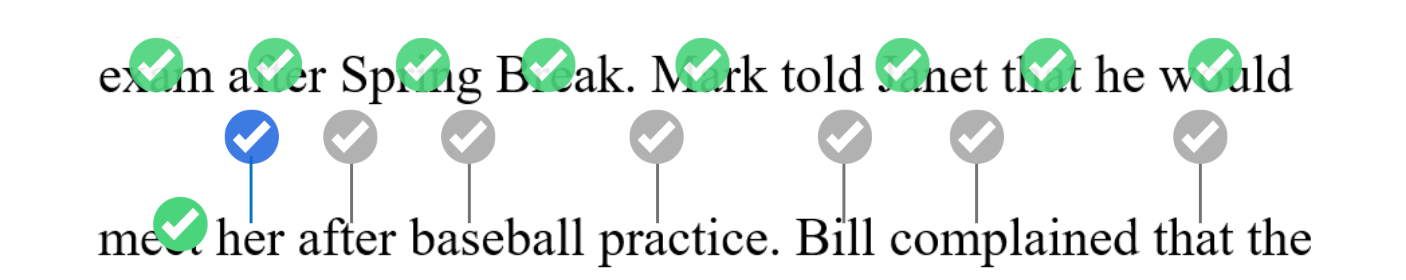}
    \caption{Suggestions are corrected based on user intervention.}
    \label{fig:situation3}
  \end{subfigure}
  
  \caption{Illustrating how user intervention through drag and drop triggers rerunning of the correction algorithm with the updated fixation position information yielding better suggestions. Green circles represent correct fixations, blue circle represents the current suggestion, and grey circles represent future suggestions.}
  \label{fig:collaboration}
\end{figure}

When the user intervenes by dragging and dropping a fixation instead of accepting the suggestion offered by the automated algorithm, the algorithm updates its internal data taking in consideration the change made by the user to make better suggestions for the remaining fixations in the same trial. This results in ideal collaboration where the automated algorithm helps the user in being faster and less subjective, while at the same time the user helps the algorithm in being more accurate.  This idea is illustrated in Figure \ref{fig:collaboration}, where a set of fixations is assigned to the wrong line in \ref{fig:situation1}, when the user drags the fixation to correct it in \ref{fig:situation2}, the correction algorithm is updated taking this intervention in consideration. Feeding the correction algorithm more accurate data yields better suggestions as illustrated in \ref{fig:situation3}.  While almost all algorithms benefit from taking in more accurate data, it is worth noting that a simple algorithm that considers the position of a single fixation at a time such as Attach, does not benefit from the updated fixation position when a user intervenes.  Therefore, the suggestions offered by Attach are not improved by the collaboration.

In combination, the four ideas give the user an ergonomic and intuitive collaborative experience with an automated correction algorithm.  We hypothesize that this combination offers a number of potential benefits: First, the process is faster than manual correction, since users mostly vet suggestions instead of manually dragging-and-dropping all of them.  Second, this process eliminates some of the subjectivity involved in correcting eye tracking data.  Third, the automated algorithm updates its correction when the user manually intervenes, which enhances the suggestions for the following fixations.  This dynamic collaboration has potential benefits to correction time, user workload, and task performance. In section \ref{sec5}, we present a usability study to assess the extent of the potential time, workload, and accuracy benefits of this approach compared to manual correction.

\section{Fix8 Features}\label{sec3}

In this section, we review the main features of Fix8 which span drift correction, eye tracking data visualization, synthetic data and distortion generation, filters, converters, and data analysis.

\subsection{Interactive Visualization}

\begin{figure}[h]
    \centering
    \includegraphics[width=0.8\textwidth]{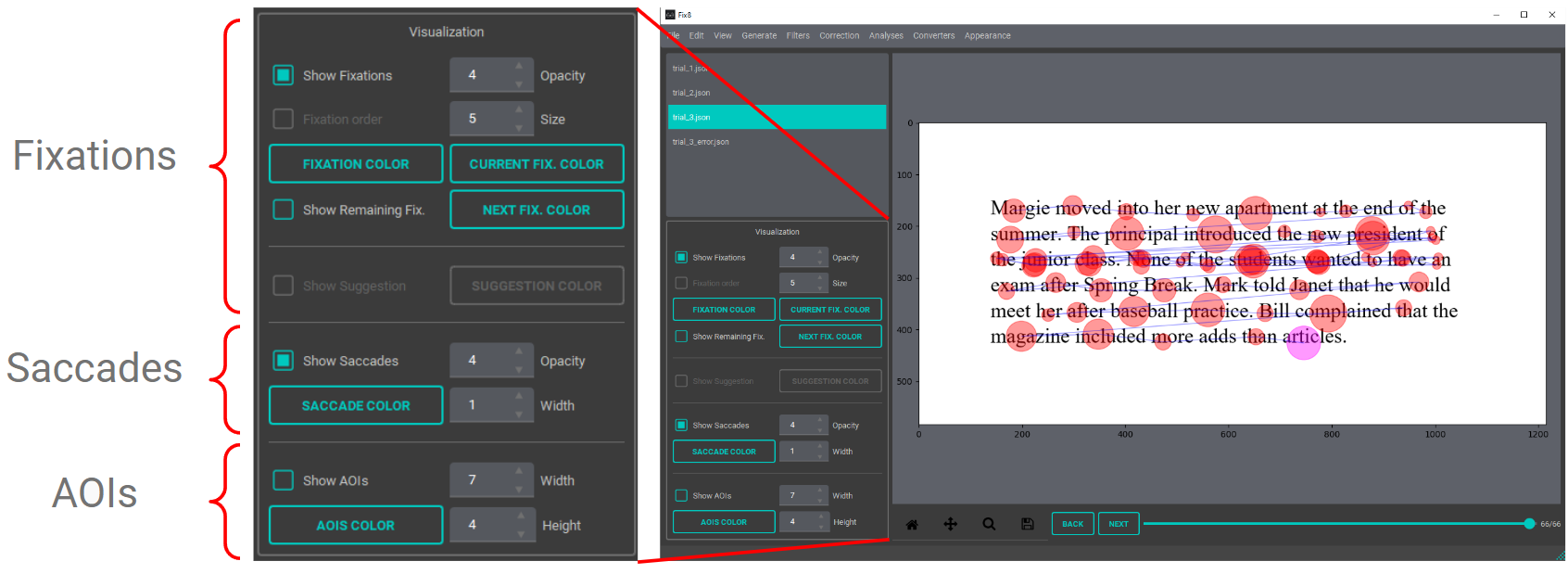}
    \caption{Visualization Panel.}
    \label{fig:visualization_panel}
\end{figure}

The visualization panel in Fix8 enables users to customize the visualization of fixations, saccades, and Areas of Interest (AOIs) to adapt to different forms of text, background colors, and research requirements. As seen in Figure \ref{fig:visualization_panel}, the visualization panel consists of three subsections for fixations, saccades, and AOIs. Each has a checkbox to control whether to display a specific visual feature or not. In the fixation subsection, users can change the overall color (default: red) of fixation, the highlighting color of the current fixation (default: magenta), and the color of remaining fixations (default: grey). 

Additionally, users can change the opacity and the relative size of the existing fixations using spin-boxes.  The size of the fixation circle is relative to the duration of the fixation.  In the assisted correction mode, users can change the color of the suggested fixation location (default: blue). In the saccades subsection, users can customize the color of saccade lines and adjust the opacity and width of saccade lines. 

\begin{figure}[h]
    \centering
    \includegraphics[width=0.7\textwidth]{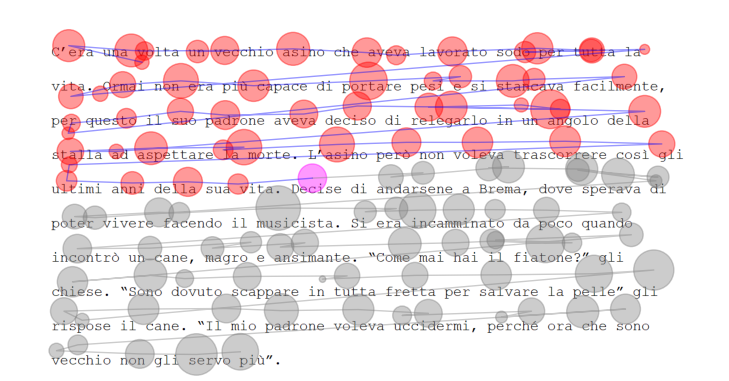}
    \caption{Visualization showing remaining fixations in grey.}
    \label{fig:remaining}
\end{figure}

In addition to previous fixation (in red) and current fixation (in magenta), users can show remaining fixations (in gray) to provide more information that might be helpful in the correction process. Figure \ref{fig:remaining} shows a visualization where remaining fixations are shown in gray.  The figure shows a real trial with data from \cite{carr2022algorithms}.  As mentioned previously, users can change these colors and adjust the transparency and relative size of fixations to enhance the visualization depending on experimental conditions and text/background colors.

\begin{figure}[h]
\centering
  \begin{subfigure}{\textwidth}
    \centering
    \includegraphics[scale=.3]{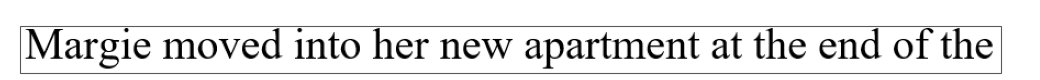}
    \caption{Line Level AOIs}
    \label{fig:line_level_aoi}
  \end{subfigure}

  \begin{subfigure}{\textwidth}
    \centering
    \includegraphics[scale=.3]{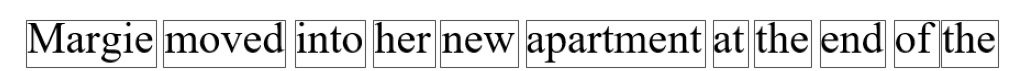}
    \caption{Word Level AOIs}
    \label{fig:word_level_aoi}
  \end{subfigure}

  \begin{subfigure}{\textwidth}
    \centering
    \includegraphics[scale=.3]{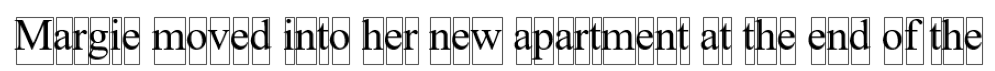}
    \caption{Letter Level AOIs}
    \label{fig:letter_level_aoi}
  \end{subfigure}

  \caption{Different Level of AOIs}
  \label{fig:aoi}
    
\end{figure}

In the AOIs subsection, users can change the color of the AOIs (default: black) and adjust the width and height thresholds for the automatically generated AOIs. Controlling AOI dimensions is particularly important for subsequent analysis on the letter level, word level,or line level. As seen in Figure \ref{fig:aoi}, Fix8 is set to the word level by default, but users have the option to adjust settings to the letter level and line level. Users can also tailor the height of AOIs to control how tight AOIs are around each line.

Fix8 can identify AOIs within a given image based on basic image processing operations. This functionality is particularly useful when precise localization of text or other significant regions is required. This is functionality is achieved by converting the image into a gray-scale black-and-White image, then this image is scanned vertically and horizontally using the height and width thresholds chosen by the user to identify transitions between black and white regions. When a space between text blocks exceeds either the height threshold or the width threshold a new area of interest is marked.

The height and width thresholds play a critical role in the precise detection of Areas of Interest within an image, significantly impacting the granularity and accuracy of the detected AOIs. Smaller height and width thresholds yield higher granularity, allowing the tool to detect finer details within the image, such as single letters. This is particularly important when the text or objects within the image are closely spaced.  On the other hand larger height and width thresholds yield lower granularity, allowing the tool to consider a complete line as a single area of interest.  We previously implemented this technique for detecting areas of interest in EMTK\footnote{https://github.com/nalmadi/EMIP-Toolkit} and the idea behind it was originally presented in EyeCode\footnote{https://github.com/synesthesiam/eyecode}.

In addition, Fix8 offers a color blind visualization under the ``Appearance" menu.  The color-blind mode is designed to enhance accessibility for users with color blindness by automatically adjusting the color of fixations and saccades to a color-blind accessible colors. Effective colorblind-friendly design involves avoiding the red-green combination \citep{Summerbell_2022}. To ensure the easy distinction, we have selected a magenta-green-blue palette.

\subsection{Generating Synthetic data and Distortions}
% Go over synthetic data generators
Fix8 offers the ability to generate synthetic eye tracking data from images of text, which could be subsequently used for testing correction algorithms, filters, or analysis metrics.  Four data generators are offered, and they range in complexity from a simple data generator that places a fixation on every word, to generators that replicate within-line and between-line regressions and skipping.  All generators rely on the AOI information calculated inside Fix8.

\begin{figure}[h]
    \centering
    \includegraphics[width=.65\textwidth]{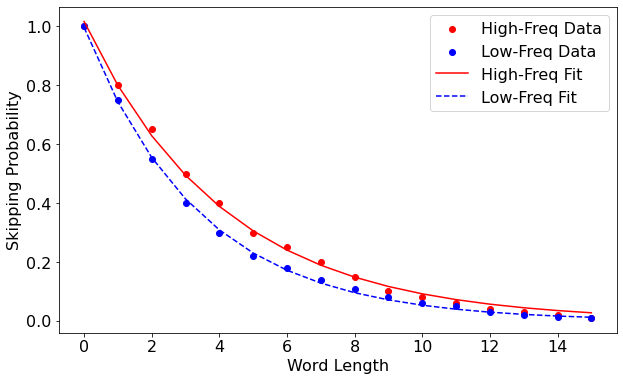}
    \caption{Skipping rate as a function of word length replicating \cite{brysbaert1998word}.}
    \label{fig:Skipping_rate}
\end{figure}

The first generator is the simplest and it places a fixation at the optimal viewing position (OVP) of each word with a random dispersion controlled by the user.  This allows the user to add a small amount of noise to the position of the fixation around the optimal viewing position.  The Second generator implements the same concept and it adds word skipping according to an exponential distribution controlled by the user.  The generator implementation replicates the exponential distribution presented by \cite{brysbaert1998word} for word skipping as a function of word length.  Fix8 allows the user to control the exponential distribution through an approximation of letter-width in pixels (to approximate the number of letters from AOI width), Lambda, and the constant K.  When the default values are used, the function replicates data from \cite{brysbaert1998word}.  Figures \ref{fig:Skipping_rate} shows the word skipping distribution used by Fix8, and Equation \ref{eq:skipping} describes the exponential distribution used to calculate skipping probability for a given word.  Word frequency is ignored by the generator in the current implementation of word skipping.

\begin{equation}
\text{skip\_probability(length)} = k \cdot e^{-\lambda \cdot \text{length}}
\label{eq:skipping}
\end{equation}

The equation describes the skip probability for a given word based on its length.  If the random number generated by a Gaussian distribution is less than the skip probability, the generator skips this word.

The third generator applies the same concept to generate data containing within-line regressions with a regression probability chosen by the user, and similarly the fourth generator applies the same concept to generate synthetic data containing between-line regressions.  To add to the realism of the generated data, all generators generate fixation duration in addition to fixation coordinates.  The synthetic duration is based on equation \ref{eq:simple_EZ}, where each word takes a minimum of 100 milliseconds and each letter in the word adds 40 milliseconds to the duration of fixation.  This results in a more realistic synthetic data where the fixation duration on each word is proportional to the length of the word, simulating the length effect found in real data \citep{rayner1998eye}.

\begin{equation}
\text{duration} = 100 + \text{len(token)} \times 40
\label{eq:simple_EZ}
\end{equation}

\begin{figure}[H]
  \begin{subfigure}{0.49\textwidth}
    \centering
    \includegraphics[width=1\linewidth]{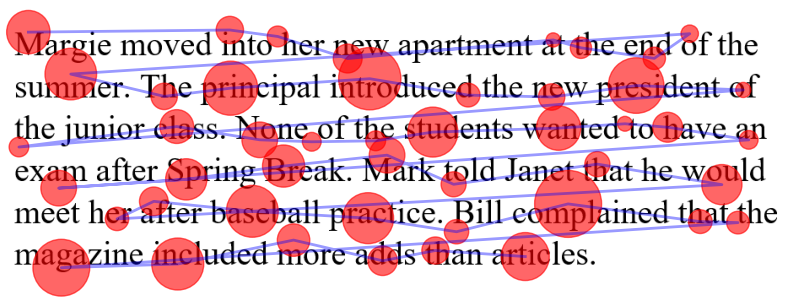}
    \caption{Noise}
    \label{fig:noise_results}
  \end{subfigure}%
  \begin{subfigure}{0.49\textwidth}\quad
    \centering
    \includegraphics[width=1\linewidth]{figs/Fix8-slope.png}
    \caption{Slope}
    \label{fig:slope_results}
  \end{subfigure}

    \medskip
    
  \begin{subfigure}{0.49\textwidth}
    \centering
    \includegraphics[width=1\linewidth]{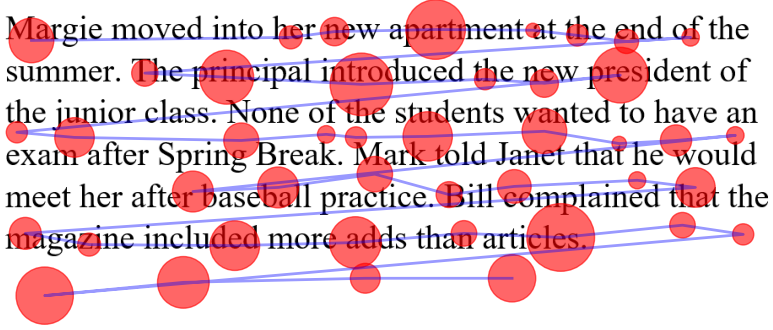}
    \caption{Shift}
    \label{fig:shift_results}
  \end{subfigure}
  \begin{subfigure}{0.49\textwidth}\quad
    \centering
    \includegraphics[width=1\linewidth]{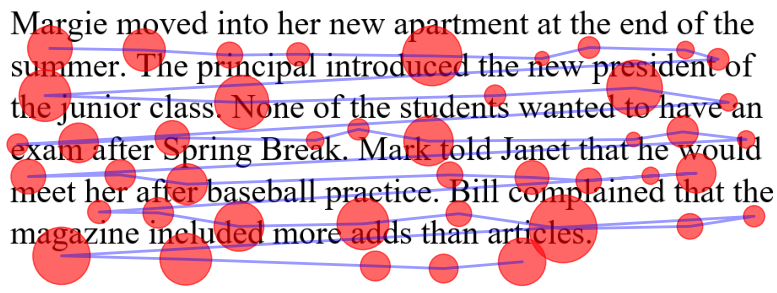}
    \caption{Offset}
    \label{fig:offset_results}
  \end{subfigure}

  \caption{Examples of distortion generators.}
  \label{fig:distortion_generators}
\end{figure}

In addition to the synthetic data generators, four distortion generators are offered to Fix8 users.  The distortion generators include noise, slope, and shift that were presented by \cite{carr2022algorithms} and offset that was presented by \cite{al2024advancing}.  As seen in Figure \ref{fig:noise_results}, noise generator introduces a distortion in the fixation coordinates according to a value chosen by the user, resulting in moving each fixation from its initial position in a random direction.  Noise distortion affects each fixation separately where one fixation might be moved up, another fixation might be moved down.  The slope distortion moves all fixations on one side of the screen simulating a systematic calibration distortion, as seen in Figure \ref{fig:slope_results}.  The distortion is controlled by the user through a distortion magnitude that determines how much the fixations are moved.  Shift distortion is similar as it pushes down fixation near the bottom of the screen while keeping fixations at the top stationary, as seen in Figure \ref{fig:shift_results}.  The amount of distortion introduced to each fixation is proportional to how far it is from the topmost fixation.  Offset distortion, applies a uniform offset to all fixations in a trial moving them up or down on the y-axis according to a magnitude determined by the user, as seen in Figure \ref{fig:offset_results}.

\subsection{Drift Correction}
Fix8 supports manual, automated, and assisted (semi-automated) correction of eye tracking data.  Through the graphical user interface, users can simply drag and drop fixations to correct them manually.  The progress bar offers an insightful view of previous and next fixations to aid the correction process.  For automated and assisted correction, Fix8 relies on algorithms that were reviewed and implemented by \cite{carr2022algorithms} and hybridizations of these algorithms \citep{al2024advancing}. Table \ref{table:algos} lists the algorithms and their description, we use the same names proposed by \cite{carr2022algorithms}.

\begin{table}[h]
\caption{The correction algorithms used in Fix8 and their description.}
\label{table:algos}
\begin{tabular}{|c|p{9cm}|}
\hline
\textbf{Algorithm} & \textbf{Description} \\ \hline
Attach  & A positional algorithm that attaches every fixation to its closet line  \citep{carr2022algorithms}.            \\ \hline
Chain    & A positional algorithm that links fixations together to form sequences of consecutive fixations, then assigns them to the line closest to the mean of their y-values \citep{schroeders2019popeye}.            \\ \hline
Cluster  & A relative positional algorithms that assigns fixations with similar y values to the same line using k-means clustering \citep{schroeders2019popeye}.           \\ \hline
Merge    & A relative positional algorithms that creates progressive sequences of consecutive fixations and merge them into larger sequences until the number of sequences equals the number of lines of text \citep{vspakov2019improving}.           \\ \hline
Regress & A relative positional algorithms that considers the fixations to be a cloud of randomly distributed points and fits regression lines to this cloud \citep{cohen2013software}.            \\ \hline
Stretch & A relative positional algorithms that determines an x-offset, y-offset, and scaling factor that minimizes alignment error between fixations and text lines \citep{lohmeier2015experimental}. \\ \hline
Segment & A Sequential algorithm that divides fixation sequence into discrete subsequences and then maps them to the lines of text from top to bottom \citep{abdulin2015person}.            \\ \hline
Slice  & A sequential algorithm that finds sequences of fixations that likely belong to the same line, then sequences are mapped to lines from top to bottom \citep{glandorf2021slice}.     \\ \hline
Warp  & A sequential algorithm that utilizes Dynamic-Time Warp to assign fixations to words minimizing the Euclidean distance between fixation positions and word centers \citep{carr2022algorithms}.     \\ \hline
Warp+Attach  & A Hybrid algorithm that splits regressions and corrects the non-regressive fixations using Warp, then adds regressions and corrects the combination with Attach \citep{al2024advancing}.     \\ \hline
Warp+Chain  & A Hybrid algorithm that splits regressions and corrects the non-regressive fixations using Warp, then adds regressions and corrects the combination with Chain \citep{al2024advancing}.     \\ \hline
Warp+Regress  & A Hybrid algorithm that splits regressions and corrects the non-regressive fixations using Warp, then adds regressions and corrects the combination with Regress \citep{al2024advancing}.     \\ \hline
Warp+Stretch  & A Hybrid algorithm that splits regressions and corrects the non-regressive fixations using Warp, then adds regressions and corrects the combination with Stretch \citep{al2024advancing}.     \\ \hline

\end{tabular}
\end{table}

Fix8 users can select an algorithm from the Correction menu to either automatically correct all fixations using the select algorithm instantly, or use assisted correction to receive a suggested correction for each fixation.  As mentioned in the previous section, the suggested correction is based on the selected automated algorithm, and it gives the user the option to accept the suggestion, correct the fixation manually, or leave the fixation where it is.  In addition, Fix8 gives the user the ability to use keyboard shortcuts in manual and assisted correction, making the correction process faster.

\subsection{Filters}
Fix8 incorporates four filters for the refinement of eye-tracking data. The filters are based on the survey by \cite{eskenazi2023best}.  The survey provides a taxonomy of the data cleaning methods and thresholds commonly used to remove eye movements that are not reflective of lexical processing.  Here we describe our implementation of the four filters in detail.

\textbf{Temporal Filters}, are ``temporal cut-offs" filters that enable the removal of fixations such that their time durations exceed or fall below certain thresholds. The greater-than filter removes fixations with durations above a certain threshold chosen by the user (\cite{eskenazi2023best} suggests 800ms as a common value), and the less-than filter removes fixations that fall bellow a threshold chosen by the user.  The less-than filter is intended to remove very short fixations that fall bellow the 80ms threshold suggested by \citep{eskenazi2023best}.

\textbf{Outlier Filter} removes fixations with durations that significantly deviate from the mean. Specifically, it enables the exclusion of fixations whose duration exceeds a user-specified threshold of standard deviations from the average fixation duration. This facilitates the removal of any outlier fixations that may be too short or too long in the data.  Such fixations are often a result of a small error or they are not reflective of cognitive processes \cite{eskenazi2023best}.

\textbf{Merge Fixations Filter} aims to correct for the over-sensitivity of eye-tracking in classifying fixations from saccades. It operates by merging very short fixations that are close to a larger fixation. The filter accepts inputs for both the maximum fixation duration threshold (for the fixations to be merged) and the max distance (in pixels, for the fixations to be merged with), enabling the user to control the merging process.

\textbf{Outside Screen Filter} works by removing any fixations that lie outside the dimensions of the stimulus on the screen.  Participants might look at the keyboard or an input device outside the dimentions of the screen, and this filter removes such fixations from the data.

\subsection{Analyses}
Utilizing the Area-Of-Interest (AOI) information, users can perform multiple levels of eye tracking data analysis. Users can generate analyses reports including fixation report, saccade report, AOI report, and duration metrics analysis as an AOI metrics report.  Fixation reports output all fixation details to a Comma-Separated Values file (CSV), and saccade reports do the same for saccades.  The AOI report performs hit-testing generate a CSV file of each fixation and the area of interest it lays over. The file includes all necessary details for subsequent analysis such as fixation coordinates, AOI coordinates, AOI dimensions, line number, word number, and stimulus image as seen in Table \ref{tab:hit_test}.

\begin{table}[h]
    \centering
    \begin{tabular}{cccccccccl}
    
           fix\_x&  fix\_y&  duration&  aoi\_x&  aoi\_y&  aoi\_width&  aoi\_height&  line&  part&image 
 \\ \hline
           168&  166&  300&  137.5&  147&  119&  44&  1&  1&stimulus.png
\\
           308&  166&  250&  262.5&  147&  112&  44&  1&  2&stimulus.png
\\
           399&  178&  200&  382.5&  147&  65&  44&  1&  3&stimulus.png

\\
    \end{tabular}
    \caption{Sample hit-test analysis report.}
    \label{tab:hit_test}
\end{table}

For the AOI metrics analysis, Fix8 can calculate the fixation count over each AOI and three duration metrics commonly used in eye tracking reading research \citep{rayner1998eye}. First-Fixation Duration (FFD): The duration of the first fixation on a word (in millisecond). Gaze Duration (GD): The duration of all the fixations on a word before moving to the next word (in millisecond). Total Time (TT): The sum of all fixation durations on a word, including fixations from regressions (in millisecond). The resulting CSV file represents an area of interest at each row, AOI details, fixation count, and the three duration metrics.

\subsection{Data Converters and Datasets}
Eye trackers record data in different file formats, and Fix8 allows for three popular eye tracking data formats and provides converters between them.  The ASCII format, often generated by EyeLink eye trackers, can be converted into Comma-Separated Value (CSV) or JSON file formats using Fix8.  Fix8 includes the ability to convert an entire ASCII experiment or a single trial instantly.  The supported CSV and JSON formats reflect two levels of detail that might be necessary depending on the type or research being conducted.  The CSV file provides all the information present in the ASCII file including timestamp, eye event (i.e. fixation/saccade/blink), x coordinate, y coordinate, duration, pupil, saccade amplitude, and peak velocity.  On the other hand, JSON files provide a simple representation of the data that consists of an ordered list of fixations, where each fixation is represented only by its x and y coordinates and duration.

\begin{lstlisting}[language=Python, caption=Fix8 JSON file format with two levels of details.]
detailed_json_data = {
    'participant': '1',
    'trial': '3',
    'time_stamps': [100, 103],
    'fixations': [[1, 2, 3], [4, 5, 6]]
}

simple_json_data = {
    'fixations': [[1, 2, 3], [4, 5, 6]]
}
\end{lstlisting}

Our two objectives are making the JSON file as simple as possible and allow users to add more details to their files without the need to modify Fix8.  Our JSON format achieves the two objectives by maintaining a dictionary where fixations are represented as a list.  If users decide to add other details to their files, \textbf{such as participant ID, trial number, or fixation timestamps} (or anything else), they are able to do so without needing to change anything in how Fix8 reads their files.  Listing 1 shows examples of the JSON file format with two levels of detail that are acceptable by Fix8.

\begin{table}[]
\caption{Fix8 includes trials from the following eye tracking datasets.}
\label{tab:datasets}
\begin{tabular}{|l|l|l|}

\hline
\multicolumn{1}{|c|}{\textbf{Dataset}} & \textbf{Eye Tracker} & \multicolumn{1}{c|}{\textbf{Description}}                                                                \\ \hline

Carr2022                               & EyeLink 1000 Plus    & \begin{tabular}[c]{@{}l@{}}33 adults and 140 children reading either of two sets of\\six passages in  Italian \citep{carr2022algorithms}.\end{tabular}     \\ \hline
EMIP2021                        & SMI RED 250          & \begin{tabular}[c]{@{}l@{}}216 programmers reading two programs in Java\\  \citep{bednarik2020emip}.\end{tabular}     \\ \hline
AlMadi2018                             & EyeLink 1000         & \begin{tabular}[c]{@{}l@{}}8 participants reading 16 single-line sentences \\ in English \citep{al2018constructing}.\end{tabular}   \\ \hline
GazeBase                               & EyeLink 1000         & \begin{tabular}[c]{@{}l@{}}322 participants reading poetry in English \\ \citep{griffith2021gazebase}.\end{tabular}     \\ \hline
MET\_Dataset                           & EyeLink 1000 Plus    & \begin{tabular}[c]{@{}l@{}}40 participants reading 97 single- and  multi-line\\ passages in  English, Spanish, Chinese, Hindi, \\Russian, Arabic, Japanese, Kazakh, Urdu, and\\ Vietnamese \citep{raymond2023dataset}.\end{tabular} \\ \hline
\end{tabular}
\end{table}

% talk about the datasets already included in Fix8
Fix8 comes with several public eye tracking datasets that cover a wide range of languages and experimental conditions.  Table \ref{tab:datasets} shows the five data sets and their details.  Each dataset brings a unique focus, whether it is comparing children and adults, analyzing the way programmers read code, or exploring reading across different languages. This variety under Fix8 further facilitates eye tracking research across various disciplines, including cognitive science, linguistics, and software engineering.

\subsection{Related Tools}
Here we compare Fix8 to existing tools that offer features for correcting eye tracking data at some level.  We thank the creators of the related tools for providing adequate documentation, most made their code publicly available, and answered our questions which we rely on in creating Table \ref{tab:tools}.

\begin{table}[h]
\centering
\caption{Comparison of correction tools and their capabilities. \Circle = not provided, \LEFTcircle = partially provided, and \CIRCLE = provided. }
\label{tab:tools}
\begin{tabular}{lllllll}
 & \rotatebox[origin=c]{70}{Fix8} & \rotatebox[origin=c]{70}{EyeDoctor} & \rotatebox[origin=c]{70}{popEye} & \rotatebox[origin=c]{70}{eyekit} & \rotatebox[origin=c]{70}{Data Viewer} & 
 \rotatebox[origin=c]{70}{EyeMap} \\ \hline

GUI                 & \CIRCLE & \CIRCLE     & \Circle  & \Circle   & \CIRCLE   & \CIRCLE  \\
Programming API     & \LEFTcircle & \Circle & \CIRCLE  & \CIRCLE  & \Circle    & \Circle \\
Manual correction   & \CIRCLE & \CIRCLE     & \CIRCLE  & \Circle   & \CIRCLE   & \CIRCLE \\
Auto correction     & \CIRCLE & \Circle     & \CIRCLE  & \CIRCLE   & \CIRCLE   & \Circle \\
Assisted            & \CIRCLE & \Circle     & \Circle  & \Circle   & \Circle   & \Circle \\
Multi-algorithm     & \CIRCLE & \Circle     & \CIRCLE  & \CIRCLE   & \Circle   & \Circle \\
Filters             & \CIRCLE & \LEFTcircle & \Circle  & \LEFTcircle & \CIRCLE & \CIRCLE \\
Multi-file formats  & \CIRCLE & \CIRCLE     & \CIRCLE  & \LEFTcircle & \Circle & \LEFTcircle \\
Synthetic data      & \CIRCLE & \Circle     & \Circle  & \Circle  & \Circle   & \Circle \\
Metrics             & \CIRCLE & \LEFTcircle & \CIRCLE  & \CIRCLE  & \CIRCLE   & \CIRCLE  \\
Open-source         & \CIRCLE & \CIRCLE     & \CIRCLE  & \CIRCLE  & \Circle   & \CIRCLE  \\
Ongoing support     & \CIRCLE & \Circle     & \CIRCLE  & \CIRCLE  & \CIRCLE   & \Circle  \\
Batch processing    & \Circle & \Circle     & \CIRCLE  & \CIRCLE  & \CIRCLE   & \Circle  \\
Stimuli creation    & \Circle & \Circle     & \Circle  & \CIRCLE  & \CIRCLE   & \CIRCLE  \\
Extensive analyses  & \Circle & \Circle     & \Circle  & \Circle  & \CIRCLE   & \Circle  \\
AOI manipulation    & \LEFTcircle & \Circle & \Circle  & \CIRCLE  & \CIRCLE   & \LEFTcircle  \\
Browser Interface    & \Circle & \Circle     & \Circle  & \Circle  & \Circle & \CIRCLE 
\end{tabular}
\end{table}

There are three tools that offer a Graphical User Interface (GUI) EyeDoctor\footnote{https://websites.umass.edu/eyelab/software/}, EyeLink Data Viewer\footnote{https://www.sr-research.com/data-viewer/}, and EyeMap \citep{tang2012eyemap}. EyeDoctor is an open-source tool designed to allow users to manipulate and clean eye movement data, and it offers manual correction along with a visualization of the data.  The tool supports data from SR-Research Eyelink II eye tracker or the Fourward Technologies DPE eye tracker.  On the other hand, EyeLink Data Viewer is free but not open-source, and it includes features for manual and automated correction.  Data Viewer is a powerful tool that implements elaborate filters and it can generate data reports and eye movement metrics.  The third GUI tools is EyeMap \citep{tang2012eyemap}, which is an open-source tool that was specifically developed for working with eye tracking data in reading research.  EyeMap offers features for visualization, AOI detection, and data analyses. The tool supports binocular data analysis across a range of writing systems and natural languages. EyeMap is unique in that it can run in a web browser.

Two tools do not offer a GUI, popEye\footnote{popEye: https://github.com/sascha2schroeder/popEye} is an open-source R package for analyzing data from different experimental paradigms \citep{schroeders2019popeye}, and it includes several automated algorithms for correcting eye tracking data in reading tasks. eyekit\footnote{eyeKit: https://jwcarr.github.io/eyekit} is an open-source Python package for analyzing eye tracking data in reading tasks, and it offers automated correction through its ``snap\_to\_lines()" function.  eyeKit supports multiple automated correction algorithms, and it implements partial filtering capabilities through the ``discard()" function.

As seen in Table \ref{tab:tools}, Fix8 includes a large array of useful features for reading research in general, and it is easily extendable as it is written entirely in Python under an open-source license.  At the same time, the table highlights some limitations including offering some but not all features in a programming API, and more importantly the limitation in running batch tasks such as automated correction of multiple trials or generating analyses reports for a group of trials at once.  Currently, Fix8 can processes one trial at a time using the GUI.  Also, Fix8 does not offer features for designing experiments or stimuli creation, although researchers could use the data generation feature to explore what data might look like for a given stimulus. The AOI manipulation offered is limited, the tool does not yet support manually drawing AOIs or editing AOIs inside the tool.  Fix8 relies on CSV files to allow for editing AOI in a program like Microsoft Excel, which could be read by Fix8 after editing.  Finally, Fix8 does not offer a browser-based interface, which remains a unique feature for EyeMap.

\section{Usability Study}\label{sec4}
In this section, we present the details of our IRB approved experiment (Colby College IRB \#2023-064) and the sources of data we rely on to assess the proposed assisted correction approach.

\subsection{Participants}
Our within-subject study relies on data from 14 participants in total. The participants were college students, five participants had a previous experience in correcting eye-tracking data and nine were novices.  All participants are above the age of 18, five are Female, and 11 are Male.  Participation was voluntary, and participants were awarded a $\$20$ gift card upon completing the experiment.

\subsection{Material}

Participants were given 36 eye tracking trials to correct, 18 trials came from real eye tracking datasets.  Four trials were from GazeBase \citep{griffith2021gazebase}, four were from the Eye Movement In Programming (EMIP) dataset \citep{bednarik2020emip}, five were from the Multilingual Eye Tracking (MET) dataset \citep{raymond2023dataset}, and five were from the Al Madi et al. \citep{al2018constructing}.  This provides a diverse set of real trials in reading English, Source Code, Hindi, Japanese, Russian, Spanish, and Urdu.  The trials also varied in the number of fixations, as some trials only contained tens of fixations while other trials had hundreds of fixations.

\begin{figure}[h]

\centering
\includegraphics[width=.32\textwidth]{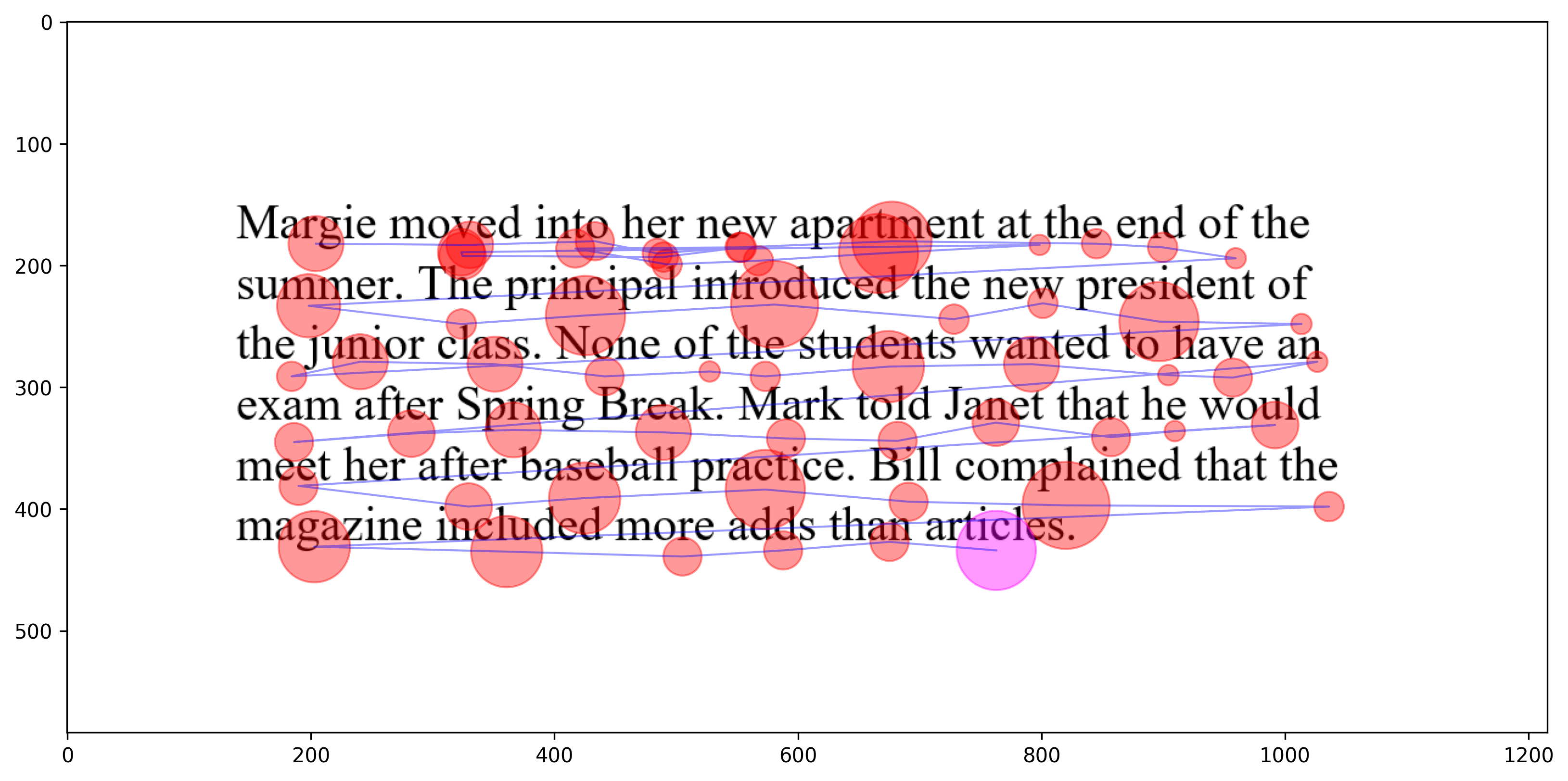}\hfill
\includegraphics[width=.32\textwidth]{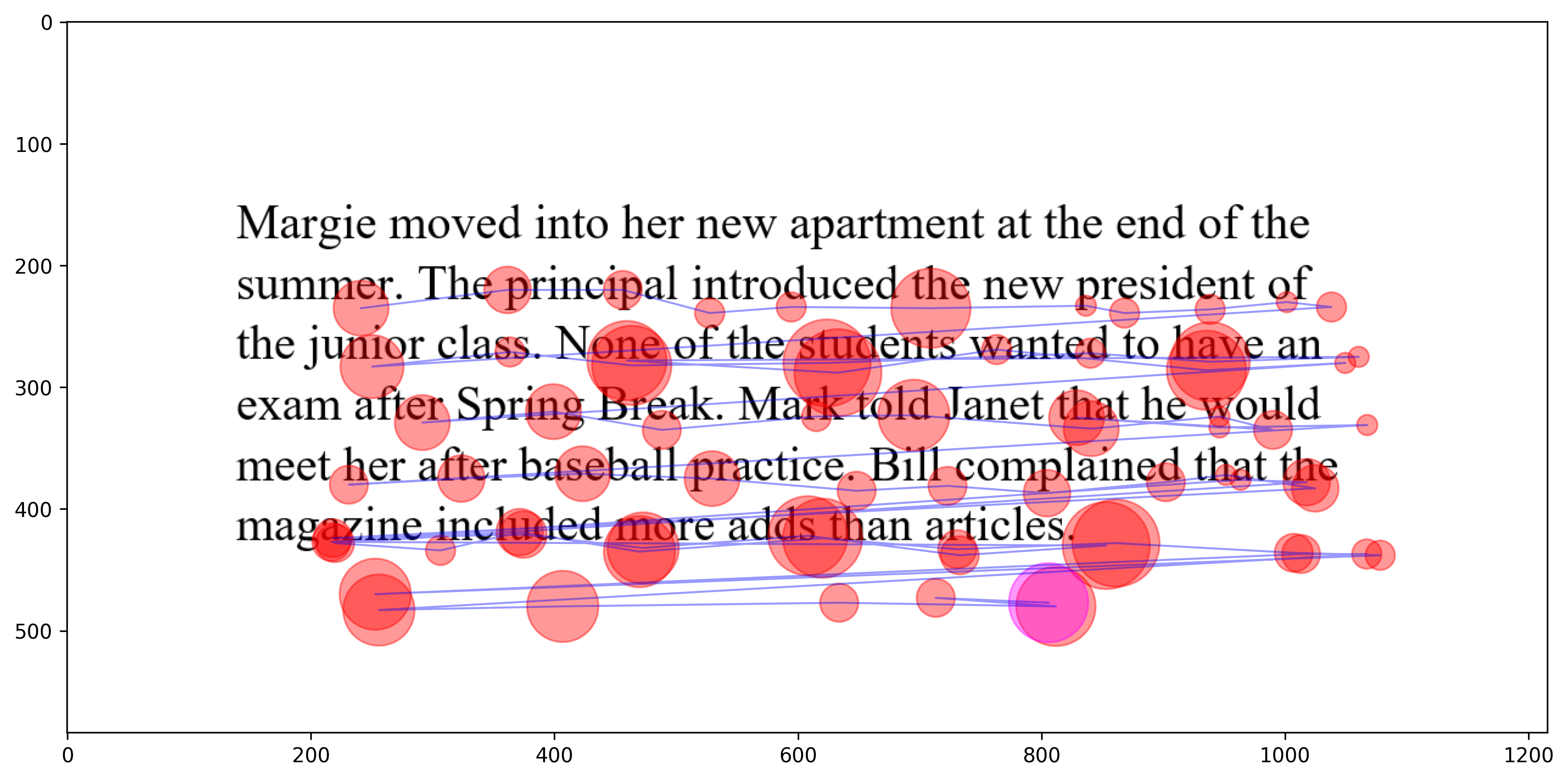}\hfill
\includegraphics[width=.32\textwidth]{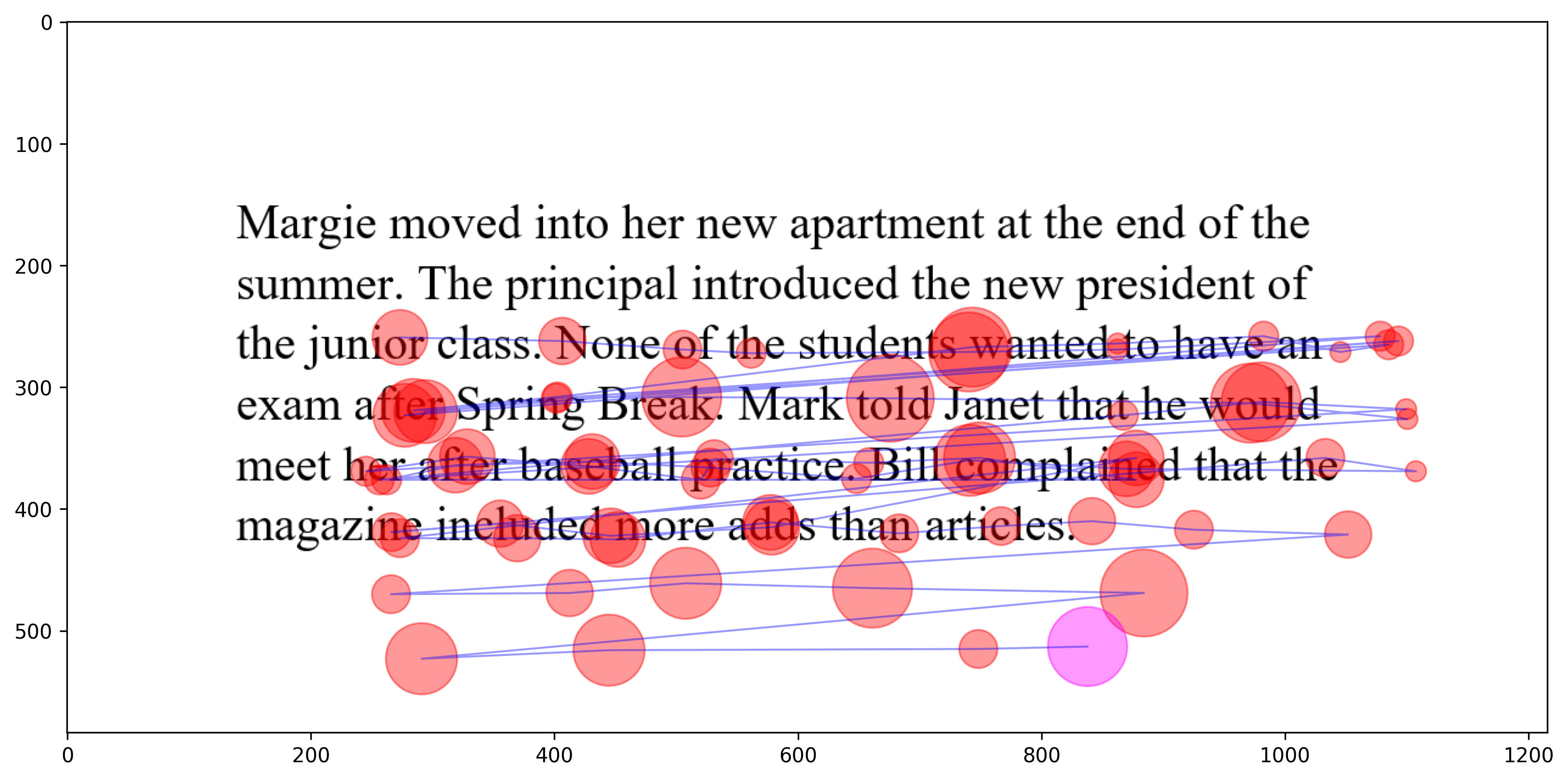}

\caption{Three synthetic trials with varying levels of offset distortion from minor (left) to extreme (right).}
\label{fig:offsets}

\end{figure}

For the purpose of accurately assessing the correction accuracy, we include in our data 18 synthetic trials with varying levels of distortion and regression, and these trials include noise, offset, shift distortions, within-line and between-line regressions.  As seen in Figure \ref{fig:offsets}, we picked synthetic trials at low, medium, and high levels of distortion for each type of distortion.

The synthetic data was generated using the same methodology mentioned in the previous section, and it contains no distortions initially.  Distortions are introduced with varying magnitudes, and the corrections made by participants are compared to the initial generated data to assess correction accuracy.  If a fixation is moved back to the initial line it was on, it is considered correct, and the accuracy of a trial is calculated as the percentage of fixations that were moved back to their initial lines.

\subsection{Procedure}
Through Zoom video conferencing, each participant was met by a member of the research team to explain the procedure and goals of the experiment and go over an electronic consent form. Regardless of experience level, participants were given a 30-minute tutorial on eye tracking data correction in reading tasks, including giving participants the opportunity to correct a test trial with supervision.  Next, the participant reports their experience in correcting eye tracking data.  Participants were asked to correct 36 eye tracking trials half using manual and half using assisted correction within a week (some participants took more time to complete the experiment).  All participants corrected an identical set of trials, and through a counterbalanced randomization we ensured that participants had approximately equal number of manual-synthetic, assisted-synthetic, manual-real, and assisted-real trials.  In addition, we randomized the order in which participants corrected the trials, and assigned a specific order to each participant to ensure that they do one trial manually followed by an assisted trial in an interleaving fashion to eliminate experience effects.

A specialized version of Fix8 was used in this experiment: this version records and timestamps every interaction with the user.  This includes correction condition, starting a trial, moving fixations, accepting suggestions, and the time taken to complete the correction of a trial.  This user interaction data is stored in a metadata file along with each correction.  This allows us to accurately calculate the time taken to correct a trial, remove long periods of inactivity when the participant takes a break, and assess how participants use the tool.  This specialized version of Fix8 had some features disabled, such as the automatic correction button.

Once the trials have been corrected, participants uploaded their corrected files and metadata to a private online storage and filled two NASA Task Load Index (NASA-TLX) forms to assess their perceived workload in manual and assisted correction. NASA-TLX is a proven workload assessment technique that has been used in many applications during the past 35 years \citep{hart2006nasa}.  The workload assessment was followed by five Likert-scale questions to collect user opinions of Fix8 and the two correction approaches.  Finally, the participant was asked to share their feedback and suggestions through an open-ended question.

\begin{figure}[h]
    \centering
    \includegraphics[width=\textwidth]{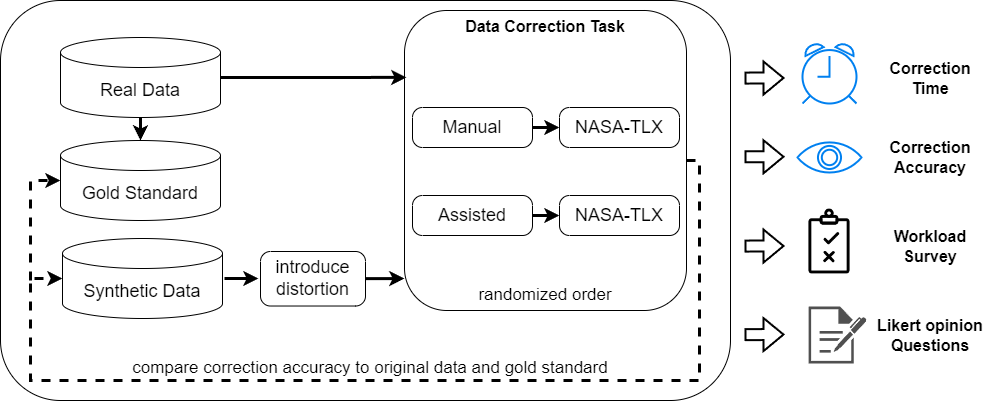}
    \caption{Overview of our experiment.}
    \label{fig:experiment}
\end{figure}

The corrected eye tracking data files and metadata files give us the ability to assess task completion time and correction accuracy.  Correction accuracy is calculated by comparing trials corrected by the participant to the original synthetic data before distortion was introduce, as illustrated by Figure \ref{fig:experiment}.  We measure accuracy as the percentage of fixations in a trial that were returned to the correct line.

To measure the accuracy with the real data, a manually corrected gold-standard set of the data is used.  The ground truth gold-standard set data comes from eight human correctors who manually corrected the real data.  Each trial was corrected by two people separately, then a software tool was used by a third person to scan and merge the data.  When the two correctors disagree, the third corrector is presented with a visualization of both corrections to make the final decision in accepting one.  A subset of this gold-standard set was presented by \citet{al2024advancing}.

The workload survey and Likert-scale questions give us the ability to assess user workload and their perceptions and preferences of the two correction methodologies.  The open-ended feedback question gives us ideas to improve the usability and add new features to Fix8.

\subsection{Results}
Here we present the results of our usability study comparing manual correction to assisted correction in Fix8.  The comparison centers around task completion time, correction accuracy, user workload, and user opinions.

\subsubsection{Completion Time and Accuracy}
A typical measure of productivity is units of work completed over time.  In our context looking at trials corrected per unit of time alone hides an important factor, which is the quality of the work.  Therefore, we examine task completion time while also looking at correction accuracy.

% time
\begin{figure}[h]
    \centering
    \includegraphics[width=0.75\textwidth]{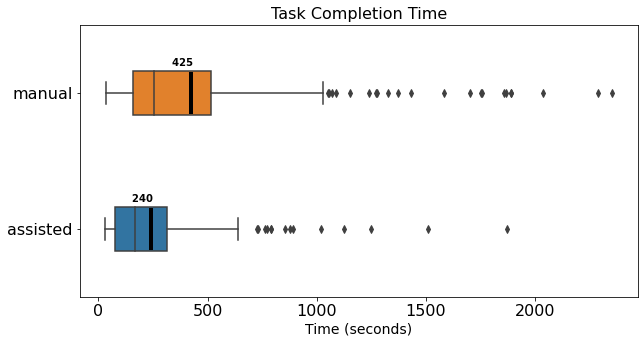}
    \caption{Task completion time (in seconds) when using manual correction compared to assisted correction.}
    \label{fig:time_results}
\end{figure}

To quantify the potential benefits of using assisted correction in comparison to manual correction, we compare the time taken to correct a trial as task completion time.  Figure \ref{fig:time_results} shows a comparison of the time taken to correct trials in manual and assisted correction approaches.  The figure shows a range of durations from 30 seconds to 30 minutes depending on the trial, as some trials contain a few fixations while some contain hundreds of fixations.  Nonetheless, the results show that on average it took 240 seconds to correct a trial using the assisted approach, while it took 426 seconds in the manual approach.  This result suggests that the assisted approach offered a 44\% time advantage per trial. A Mann-Whitney U test was employed to compare task completion time between manual and assisted correction. The analysis revealed a significant difference between the two conditions (\textit{U} = 38529.5, p $<$ .001). The effect size (Cohen's d) was calculated to be $0.52$, indicating a medium-sized effect.

% accuracy synthetic
\begin{figure}[h]
    \centering
    \includegraphics[width=0.75\textwidth]{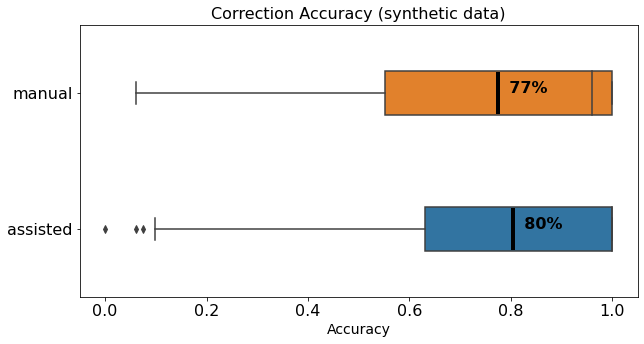}
    \caption{Correction accuracy of synthetic data with manual correction compared to assisted correction.}
    \label{fig:accuracy_results}
\end{figure}

Task completion time seems significantly shorter with assisted correction, yet correction accuracy is an important factor in our context.  Therefore, we examine the correction accuracy in correcting \textbf{synthetic data} that was corrected using the manual and assisted techniques. Figure \ref{fig:accuracy_results} shows the accuracy of manual and assisted approaches, and the values range from zero to one.  Some synthetic trials had extreme errors that were really difficult to correct, therefore, 10 trials had an accuracy score bellow 0.1.  Nonetheless, the results show that the average accuracy of manual and assisted correction are comparable.  A Mann-Whitney U test was utilized to compare correction accuracy between manual and assisted correction. The analysis revealed no significant difference between the two conditions (\textit{U}=6221.5, \textit{p}=.12). The effect size (Cohen's d) was calculated to be -0.10.

% accuracy real
\begin{figure}[h]
    \centering
    \includegraphics[width=0.75\textwidth]{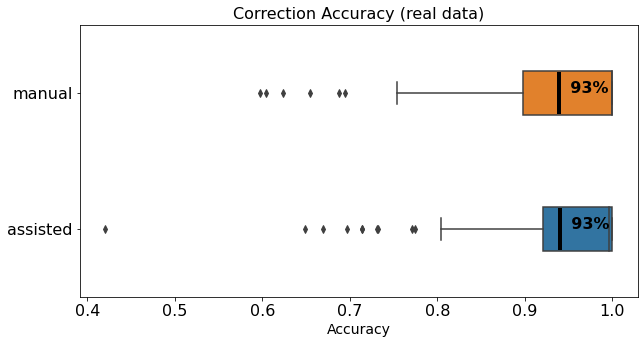}
    \caption{Correction accuracy of real data with manual correction compared to assisted correction.}
    \label{fig:accuracy_results_real}
\end{figure}

Examining the correction accuracy of \textbf{real data}, Figure \ref{fig:accuracy_results_real} shows the accuracy of manual and assisted approaches. The results show that the average accuracy of manual and assisted correction is also comparable in real data with an average accuracy of 93\%.  A Mann-Whitney U test was utilized to compare correction accuracy between manual and assisted correction. The analysis revealed no significant difference between the two conditions (\textit{U}=5807.5, \textit{p}=.43). The effect size (Cohen's d) was calculated to be -0.14.  Therefore, our task completion time and accuracy results suggest that assisted correction was significantly faster than manual correction without sacrificing accuracy.

\subsubsection{User Workload and Opinions}
Here, we focus on perceived workload when using our assisted approach in comparison to manual correction.  User workload perceptions and opinions were reported after participants have completed both manual and assisted correction.  High workload can negatively impact user performance in correcting eye tracking data, and workload and opinion assessment can provide insights into aspects of the proposed tool that may be causing cognitive or physical strain on users.

% TLX
\begin{figure}[h]
    \centering
    \includegraphics[width=.8\textwidth]{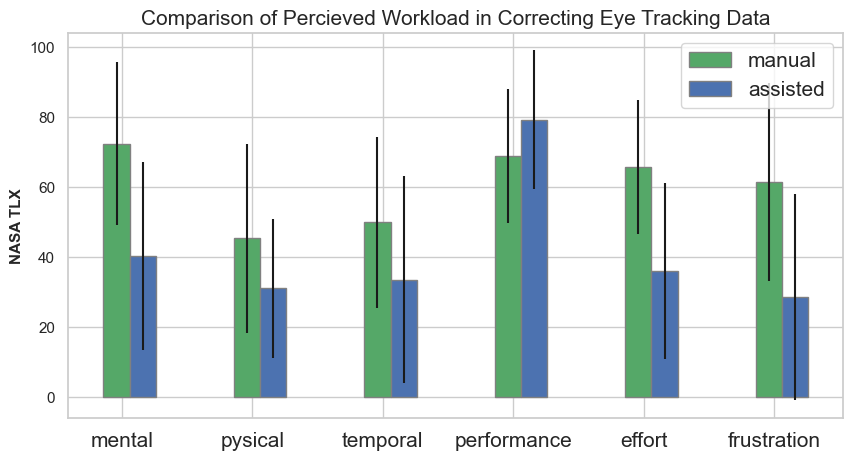}
    \caption{Comparing NASA-TLX components in manual and assisted correction.}
    \label{fig:tlx_results}
\end{figure}

A comparison of the perceived workload components in manual and assisted correction is presented in Figure \ref{fig:tlx_results}.  The NASA-TLX components show that participants preferred the assisted approach over manual correction.  This preference is reflected in reporting higher mental demand, physical demand, temporal demand, effort, and frustration with manual correction.  Moreover, users' perception of their performance was higher with the assisted approach compared to manual correction.

\begin{figure}[h]
    \centering
    \includegraphics[width=1 \textwidth]{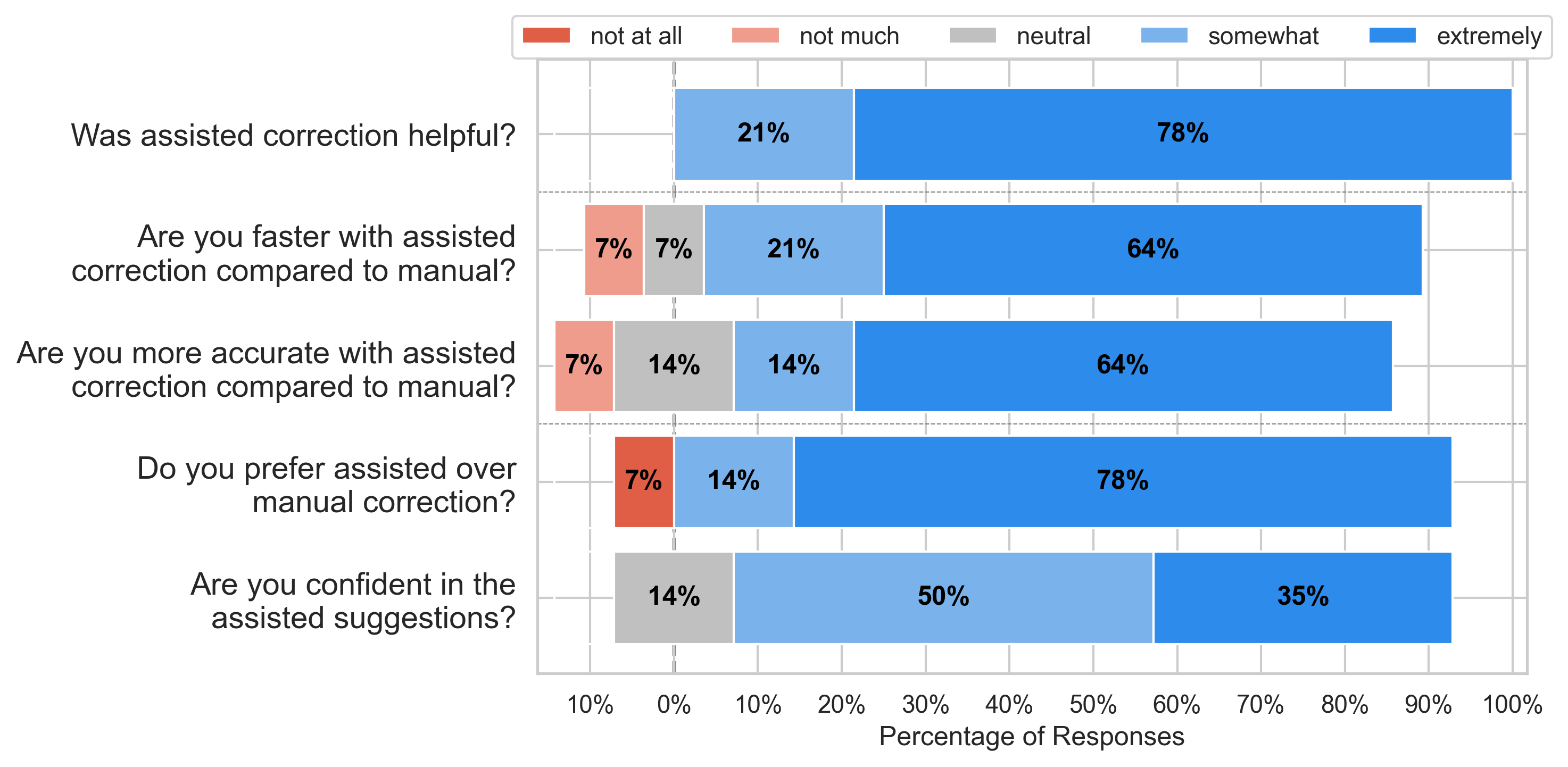}
    \caption{Summary of Likert scale survey results as reported be participants.}
    \label{fig:likert_coding}
\end{figure}

To gain a better understanding of user opinions, we asked study participants five Likert-scale questions after the experiment.  Figure \ref{fig:likert_coding} shows the results of the Likert-scale survey. Overall the results show that users had a preference and positive opinions of assisted correction over manual correction.  100\% of participants found assisted correction somewhat or extremely helpful, and 85\% of participants found assisted correction somewhat or extremely faster.  Moreover, 78\% found assisted correction more accurate, and 92\% preferred assisted correction over manual correction.  Finally, 85\% of participants were somewhat or extremely confident in assisted correction suggestions.  

In regards to open-ended feedback, participants were generous in making suggestions including several ideas that we incorporate in the current version of Fix8.  The color-blind assist idea came from one participant, also the responsiveness of the interface when users drag-and-drop fixations has been improved.

\subsection{Summary}
The results of our usability study show that the proposed assisted approach for correcting eye tracking data was significantly faster than manual correction.  On average assisted correction was 44\% faster than manual correction.  Moreover, the gain in task completion time did not result in a negative effect on accuracy, as results show that assisted correction was very comparable in accuracy to manual correction.  This is important considering that manual correction is often used as the ground truth in eye tracking data correction.  In addition, our results show that participants had a reduced workload and better perception of their performance when using the assisted approach, and generally participants had positive opinions of the assisted approach in terms of helpfulness, speed, accuracy, and confidence in its suggestions.

\subsection{Workflow Improvements}
From the results of the usability study, participant feedback, and suggestions from reviewers we made two improvements to the correction workflow that influence both manual and assisted correction.  We present the details of the two ideas that enhance the ergonomics of the data correction workflow.

\begin{figure}[h]
    \centering
    \includegraphics[width=1 \textwidth]{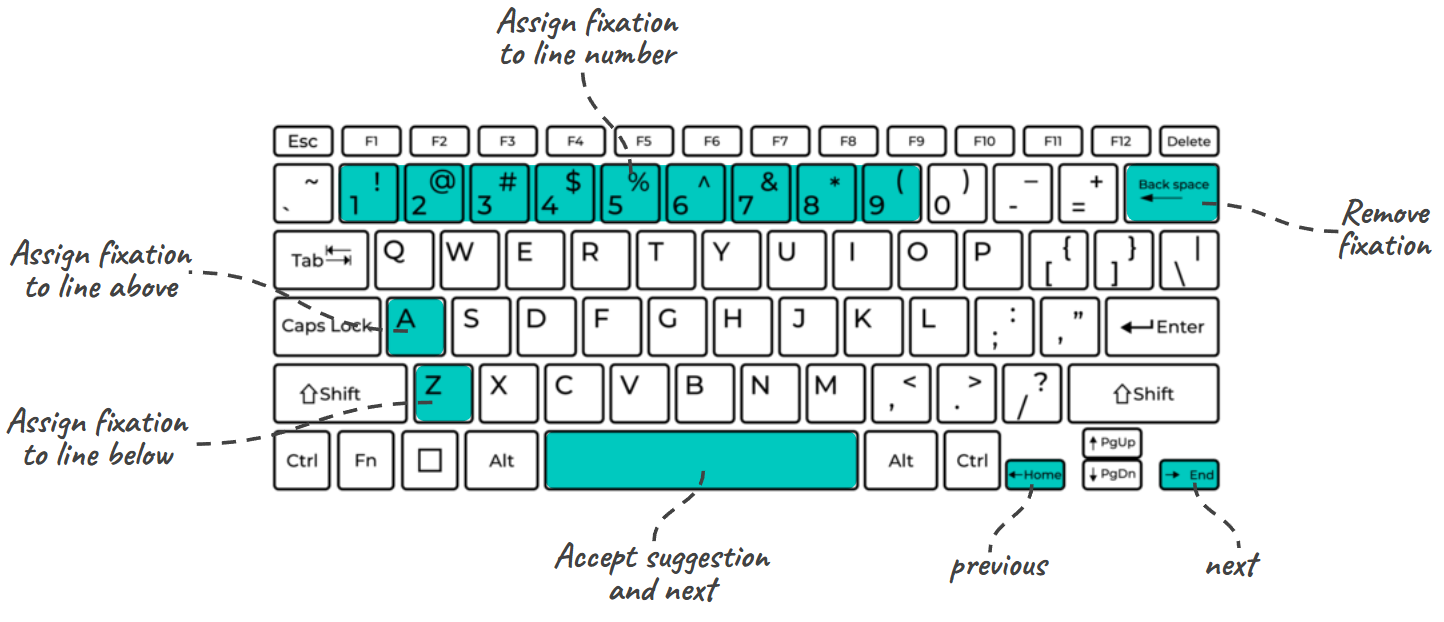}
    \caption{A keyboard shortcut layout for better ergonomics in data correction.}
    \label{fig:key}
\end{figure}

Both ideas rely on the fact that manual and assisted correction primarily involve line assignment on the y-axis, where it is less frequent that corrector make an adjustment on the horizontal position of the fixation.  Therefore, the first idea incorporates using two keys on the keyboard, the key ``a" assigns the current fixation to the line immediately above the fixation, and the key ``z" assigns the fixation the line immediately below the fixation.  Both shortcuts apply the correction and move to the next fixation, making the entire process faster and involves less effort compared to dragging-and-dropping the fixation.

The second idea is credited to reviewer Jon Carr, and it applies the same concept by utilizing the 1 to 9 keys on the keyboard.  When the number 1 is pressed, the fixation is moved to the first line in the text, and each key moves the current fixation to the line matching its number.  Although this approach is limited to trials with less than 10 lines of text, it speeds up the correction process substantially.

\section{Discussion}\label{sec5} 
The main contribution this paper makes is in presenting a semi-automated, assisted approach for correcting eye tracking data in reading tasks.  This approach is motivated by the gap between automated correction algorithms and manual correction, where manual correction remain more accurate yet time consuming, subjective, and laborious.  The presented assisted approach leverages recent advancements in automated correction algorithms with an interface that grants an easy workflow that achieves fast and accurate correction process.  Our study shows that the time benefits of using the proposed assisted approach are substantial (44\% faster on average), in addition the correction accuracy matches manual correction.  The proposed approach fills a gap between manual and automated correction by offering an approach that is faster than manual correction yet matches its accuracy.

In our usability study, the synthetic data accuracy of manual and automated correction averaged around 78\%, which is not very high compared to the expected accuracy of manual correction.  We believe that this is due to the excessive magnitude of error that some of the synthetic trials had in our experiment, where some trials were beyond what we would typically observe in a real eye tracking experiment. This belief is affirmed by the real data accuracy results which averaged around 93\%, which is more inline with what is expected in manual correction.

At the same time, the accuracy results challenge some of our existing notion of manual correction as the ground truth, as often presumed in eye tracking research.  The process of correcting eye tracking data whether automatic, assisted, or manual is based on informed guesses and heuristics that can be incorrect in some situations. This calls for future research to empirically assess the accuracy of manual correction and the possible factors that influence manual correction accuracy.  With its metadata logging feature, Fix8 offers the necessary infrastructure to conduct this kind of research.

The assisted correction approach combines visualization, workflow, ergonomics, and collaboration to achieve an enhanced correction experience.  Although it is difficult to isolate the role of ergonomics (i.e. keyboard and mouse) for direct comparison with manual correction in our experiment, it is clear that the proposed assisted approach is better than standard manual approaches.

The proposed assisted approach is an example of seamless collaboration between a human expert and a problem solving algorithm. With every intervention the user makes, the algorithm can update its suggestions, taking that intervention into account.  This offers an improvement in accuracy over the use of automated corrections without the laborious workload of manual correction. This reduction in the required mental and physical work might explain why users preferred the assisted workflow and reported a lower workload and higher perceived performance compared to manual correction.

Beyond the assisted correction technique, Fix8 offers a range of useful features for eye tracking data generation, visualization, filtering, and analysis in an easy and intuitive interface.  Additionally, the data converters offered allow for different data formats and eye trackers to be used, with easy extensibility as an open-source tool. It is likely that automated correction techniques will match and possibly outperform human correctors in the future, until then the proposed correction technique and features represent a combination of benefits that make Fix8 useful in eye tracking correction and eye tracking research in general.  We see Fix8 as one piece of a larger effort for building a fully open-source research-grade eye tracking infrastructure that covers experiment design, data collection, and data cleaning and analysis.
\section{Conclusion}\label{sec6}
In this paper, we presented an open source, GUI tool that offers a range of features for eye tracking data in reading research named Fix8.  This tool implements a novel semi-automated correction technique that offers faster workflow than manual correction without sacrificing correction accuracy.  This approach utilizes automated correction algorithm to offer users suggestions in the correction process, and Fix8 allows users to accept/reject suggestions, or manually correct fixations through an intuitive interface.  Fix8 also offers fully manual and fully automated correction options.

Beyond eye tracking data correction, Fix8 offers useful features for generating synthetic eye tracking data, visualization, filters, data converters, and eye movement analyses.  Through a usability study, we found that the assisted approach was faster than manual correction without any sacrifice in correction accuracy, and users reported lower workload and higher perceived performance when using the proposed approach.  Until automated algorithms outperform human correction, the proposed assisted approach remains a useful tool that combines speed with accuracy. We aim to extend the features of Fix8 through community driven feedback and participation to make eye tracking research more accessible.

\backmatter

\bmhead{Acknowledgments}
The authors would like to thank our study participants and field experts who provided their valued feedback on our work. Also, thanks to the reviewers and editorial board for providing valued comments and suggestions.

\section*{Declarations}

\bmhead{Funding} This research did not receive any grants from funding agencies in the public, commercial, or non-profit sectors.

\bmhead{Conflict of interest} The authors do not have any financial or non-financial conflict of interest to report.

\bmhead{Ethics approval} Approval was obtained from the Institutional Review Board of Colby College (IRB \#2023-064). The procedures used in this study adhere to the principles of the Declaration of Helsinki.

\bmhead{Consent to participate} Informed consent was obtained from all individual participants in the study.

\bmhead{Consent for publication}  All authors consent to publish this research.

\bmhead{Availability of data and materials} A replication package of the experiment in this paper including code and data can be found at this Open Science Framework link: \url{https://osf.io/umkp4/}

\bmhead{Code availability}  Fix8 is available as an open source tool with all the code and a collection of datasets at the following GitHub repository: \url{https://github.com/nalmadi/fix8}

A video covering the main features of the tool is available at the following link: \url{https://fix8-eye-tracking.github.io/video}

\bmhead{Open practices statement} This study did not preregister its hypotheses, methods, or analysis plans. However, the data and materials used in this research are available as an open source tool and a replication package containing all code and data.

\bmhead{Open Access} This article is licensed under a Creative Commons Attribution 4.0 International License, which permits use, sharing, adaptation, distribution and reproduction in any medium or format, as long as you give appropriate credit to the original author(s) and the source, provide a link to the Creative Commons licence, and indicate if changes were made. The images or other third party material in this article are included in the article's Creative Commons licence, unless indicated otherwise in a credit line to the material. If material is not included in the article's Creative Commons licence and your intended use is not permitted by statutory regulation or exceeds the permitted use, you will need to obtain permission directly from the copyright holder. To view a copy of this licence, visit \url{http://creativecommons.org/licenses/by/4.0/}.

\bibliography{sn-bibliography}% common bib file

\end{document}